\documentclass[preprint,aps,groupedaddress,showpacs]{revtex4}
\usepackage{amsfonts}
\usepackage{amssymb}
\usepackage{graphicx}
\usepackage{amsmath}
\usepackage{bm}
\usepackage{relsize}
\RequirePackage{xspace}
\begin{document}
\title{\mbox{}\\[10pt]
Search for excited charmonium states in $e^+e^-$ annihilation at
$\sqrt{s}=10.6$ GeV}

\author{Kui-Yong Liu$~^{(a,b)}$, Zhi-Guo He$~^{(a)}$,
and Kuang-Ta Chao$~^{(a,c)}$} \affiliation{ {\footnotesize
(a)~Department of Physics, Peking
University, Beijing 100871, China}\\
{\footnotesize (b)~Department of Physics, Liaoning University,
Shenyang 110036, China}\\   {\footnotesize (c)~China Center of
Advanced Science and Technology (World Laboratory), Beijing 100080,
China}}

\begin{abstract}
We suggest searching for excited charmonium states in $e^+e^-$
annihilation via double charmonium production at $\sqrt{s}=10.6$ GeV
with $B$ factories, based on a more complete leading order
calculation including both QCD and QED contributions for various
processes. In particular, for the C=+ states, the $\chi_{c0}(nP)$
(n=2,3) and $\eta_c(mS)$ (m=3,4) may have appreciable potentials to
be observed; while for the C=- states, the $\eta_ch_c$ production
and especially the $\chi_{c1}h_c$ production might provide
opportunities for observing the $h_c$ with higher statistics in the
future. A brief discussion for the X(3940) observed in the double
charmonium production is included.
\end{abstract}


\maketitle

\section{Introduction}
Charmonium spectroscopy has become a challenging topic in hadron
physics and QCD, because of the recent findings of possible new
charmonium states (for recent experimental and theoretical reviews
and discussions, see e.g. \cite{skwarnicki, olsen, quigg, close} and
references therein). Among others, for the puzzling state X(3872),
possible assignments of e.g.  the $2^3P_1$ and $1^2D_1$ charmonium
states and the charm-molecule have been suggested (see,
e.g.\cite{swanson06} for a comprehensive review), and it will be
helpful to search for those states in other experiments and to
clarify these assignments; The measured mass splitting between
$\psi(2S)$ and $\eta_c(2S)$ is about 50 MeV, which is smaller than
some theoretical predictions, and it is certainly useful to search
for the $\eta_c(3S)$ to see what will be the mass splitting between
the $\psi(3S)$, which could be the observed $\psi(4040)$, and the
$\eta_c(3S)$. This may be particularly interesting since according
to some potential model calculations the $\eta_c(3S)$ could lie
above 4040~MeV (see, e.g. in \cite{godfrey} the mass of $\eta_c(3S)$
is predicted to be 4060~MeV).  And the $\psi(3S)$  mass could
actually be lowered by coupling to the nearby $D^*\bar D^*$ decay
channels (note that the energy level spacing between
$\psi(3S)=\psi(4040)$ and $\psi(2S)=\psi(3686)$ is smaller than that
between $\psi(4S)=\psi(4414)$ and $\psi(3S)=\psi(4040)$, which is in
contradiction with potential model calculations unless the coupled
channel effects are considered or the assignments for
$\psi(3S)=\psi(4040)$ and $\psi(4S)=\psi(4414)$ are incorrect. The
mass spectrum of excited charmonium states will certainly provide
important information on interquark forces and color confinement. In
addition, studies of the decay and production of these states will
also be very important in understanding the underlying theory of
strong interaction --perturbative and nonperturbative QCD in view of
many outstanding puzzles in charmonium physics.

$B$ meson decays have proven to be very useful processes to find new
charmonium states. Aside from the $B$ meson decay, $e^+e^-$
annihilation at $\sqrt{s}=10.6$ GeV could also be a very useful
process in finding the excited charmonium states, since the recent
Belle experiments \cite{belle1,belle2} have found unusually strong
signals for the double charmonuim production from the $e^+e^-$
continuum, e.g., $e^+e^-\rightarrow J/\psi\eta_c, J/\psi\chi_{c0}$,
$J/\psi\eta_c(2S)$ and $e^+e^-\rightarrow \psi(2S)\eta_c$,
$\psi(2S)\chi_{c0},  \psi(2S)\eta_c(2S)$. Theoretically, the
calculated cross sections for these processes based on the leading
order Non-Relativistic QCD(NRQCD) (or more generally perturbative
QCD (pQCD)) are about an order of magnitude smaller than the
experiments \cite{bl, liu, h}. This is a big issue in charmonium
physics and NRQCD, and it still remains to be further clarified
though many considerations are suggested to understand the large
production rates in both exclusive and inclusive charmonium
production via double charm pairs in $e^+e^-$ annihilation \cite{cc}
(the theoretical predictions for the inclusive $J/\psi c \bar c$
production cross section with the color-singlet \cite{cho,yqc,ko} as
well as color-octet\cite{lhc} contributions are also much smaller
than the Belle data). Despite of these disagreements, however, we
find that the calculated relative rates of the double charmonium
production processes are roughly compatible with the Belle data
(e.g. the production cross sections of $\eta_c$, $\eta_c(2S)$, and
$\chi_{c0}$ associated with $J/\psi$ and $\psi(2S)$ are much larger
than that of $\chi_{c1}$ and $\chi_{c2}$). So, we may use the same
method as in our previous work to calculate the production rates for
the excited charmonium states in $e^+e^-$ annihilation into double
charmonia, but mainly pay attention to the relative rates for these
production processes. We hope the calculation will make sense in
predicting the relative production rates for those excited
charmonium states, and can be tested by experiments. This will be
useful not only in the search for those excited charmonium states,
but also in understanding the production mechanism itself. If the
predicted relative production rates turn out to be consistent with
experiments, it is likely that the NRQCD factorization treatment for
these processes probably still makes sense and only an overall
enhancement factor is needed and should be clarified in further
theoretical considerations (including QCD radiative corrections,
relativistic corrections, and other nonperturbative QCD effects). In
the last section we will have a discussion on recent developments in
this regard. In the following, we will calculate the leading order
production cross sections for various excited charmonium states in
$e^+e^-$ annihilation at $\sqrt{s}=10.6$ GeV in the same way as in
\cite{liu}.

\section{Formulas and calculations}
Following the NRQCD factorization formalism\cite{bbl}, the
scattering amplitude of double charmonia production can be described
as
\begin{eqnarray}
\label{amp2}   &&\hspace{-2.0cm}{\cal A}(a+b\rightarrow
Q\bar{Q}({}^{2S_\psi+1}L_{J_\psi})(p_3)+Q\bar{Q}({}^{2S
+1}L_{J})(p_4))= \sqrt{C_{L_\psi}}\sqrt{C_L}\sum\limits_{L_{\psi
z} S_{\psi z} }\sum\limits_{s_1s_2 }\sum\limits_{jk}
\sum\limits_{L_z S_z }\sum\limits_{s_3
s_4}\sum\limits_{il}\nonumber\\
&\times&\langle s_1;s_2\mid S_\psi S_{\psi z}\rangle \langle
L_\psi L_{\psi z };S_\psi S_{\psi z}\mid J_\psi J_{\psi
z}\rangle\langle 3j;\bar{3}k\mid
1\rangle\nonumber\\&\times&\langle s_3;s_4\mid S S_z\rangle\langle
L L_z ;S S_z\mid
J J_z\rangle\langle 3l;\bar{3}i\mid 1\rangle\nonumber\\
 &\times&\left\{
\begin{array}{ll}
{\cal A}(a+b\rightarrow
 Q_j(\frac{p_3}{2})+\bar{Q}_k(\frac{p_3}{2})+
 Q_l(\frac{p_4}{2})+\bar{Q}_i(\frac{p_4}{2}))&(L=S),\\
\epsilon^*_{\alpha}(L_Z) {\cal A}^\alpha(a+b\rightarrow
 Q_j(\frac{p_3}{2})+\bar{Q}_k(\frac{p_3}{2})+
 Q_l(\frac{p_4}{2})+\bar{Q}_i(\frac{p_4}{2}))
&(L=P),\\
\end{array}
\right.\nonumber\\
\end{eqnarray}
where $\langle 3j;\bar{3}k\mid 1\rangle =\delta_{jk}/\sqrt{N_c}$~,
$\langle 3l;\bar{3}i\mid 1\rangle=\delta_{li}/\sqrt{N_c}$~,~
$\langle s_1;s_2\mid S_\psi S_{\psi z}\rangle$~,~$\langle
s_3;s_4\mid S S_z\rangle$~,~ $\langle L_\psi L_{\psi z };S_\psi
S_{\psi z}\mid J_\psi J_{\psi z}\rangle$ and $ \langle L L_z ;S
S_z\mid J J_z\rangle$ are respectively the color-SU(3),
spin-SU(2), and angular momentum Clebsch-Gordon coefficients for
$Q\bar{Q}$ pairs projecting out appropriate bound states. ${\cal
A}(a+b\rightarrow Q_j(\frac{p_3}{2})+\bar{Q}_k(\frac{p_3}{2})+
 Q_l(\frac{p_4}{2})+\bar{Q}_i(\frac{p_4}{2}))$ is the scattering
 amplitude for double $Q\bar{Q}$ production and ${\cal A}^\alpha$
 is the derivative of the amplitude with respect to the relative
 momentum between the quark and anti-quark in the
bound state. The coefficients $C_{L_\psi}$ and $C_L$ can be
related to the radial wave function of the bound states or its
derivative with respect to the relative spacing as
\begin{equation}
\label{cs} C_s=\frac{1}{4\pi}|R_s (0)|^2,\ \ \
C_p=\frac{3}{4\pi}|R_p'(0)|^2.
\end{equation}
We introduce the spin projection operators $P_{SS_z}(p,q)$
as\cite{cho2,pro}
\begin{equation}
P_{SS_z}(p,q)\equiv\sum\limits_{s_1s_2 }\langle
s_1;s_2|SS_z\rangle
v(\frac{p}{2}-q;s_1)\bar{u}(\frac{p}{2}+q;s_2).
\end{equation}
Expanding $P_{SS_z}(P,q)$ in terms of the relative momentum $q$,
we get the projection operators and their derivatives, which will
be used in our calculation, as follows
\begin{equation}
\label{pjs} P_{1S_z}(p,0)=\frac{1}{2\sqrt{2}}\ \epsilon\!\!
/^*(S_z)(\not{p}+2m_c),
\end{equation}

\begin{equation}
\label{petc} P_{00}(p,0)=\frac{1}{2\sqrt{2}}\gamma_5(p\!\!\!\!
/+2m_c),
\end{equation}

\begin{equation}
\label{petca}
P_{00}^{\alpha}(p,0)=\frac{1}{2\sqrt{2}m_c}\gamma^{\alpha}\gamma_5p\!\!\!\!
/,
\end{equation}
\begin{equation}
\label{der} P_{1S_z}^{\alpha}(p,0)=\frac{1}{4\sqrt{2}m_c}
[\gamma^{\alpha}\not{\epsilon}^*(S_z)(\not{p}+2m_c)-
(\not{p}-2m_c)\not{\epsilon}(S_z)\gamma^{\alpha}].
\end{equation}

We then get the following expressions and numerical results for
various processes of double charmonium production in  $e^+e^-$
annihilation at $\sqrt{s}=10.6$ GeV. In the calculation of the short
distance coefficients, the quark and anti-quark are all on mass
shell, and the meson masses are taken to be $m_3=m_4=2m_c$. The
input parameters are $\sqrt{s}=10.6{\rm GeV},~~m_c=1.5{\rm GeV},~
\alpha_s=0.26$ (corresponding to $\mu=2m_c$ and $
\Lambda^{(4)}_{\overline{MS}}$=338~MeV), and the wave functions at
the origin are taken from a potential model calculation (see the QCD
(BT(Buchm\"{u}ller-Tye)) model in Ref.\cite{wf}): $|R_{1S}(0)|^2
=0.810{\rm GeV}^3$, $|R_{2S}(0)|^2 =0.529{\rm GeV}^3$,
$|R_{3S}(0)|^2 =0.455{\rm GeV}^3$, $|R_{1P}'(0)|^2=0.075 {\rm
GeV^5}$, $|R_{2P}'(0)|^2=0.102 {\rm GeV^5}$, and
$|R_{1D}''(0)|^2=0.015 {\rm GeV^7}$.
\subsection{The $\psi(nS)\eta_c(mS)$ production}

In the two S-wave (nS and mS) case, the
cross section for $e^++e^-\rightarrow \gamma^{*} \rightarrow
H_1(nS)+H_2(mS)$ is given by
\begin{eqnarray}
\label{jsetc} &&\sigma(e^+(p_1)+e^-(p_2)\rightarrow
H_1(p_3)+H_2(p_4))=\nonumber\\
&&\frac{2\pi\alpha^2\alpha_s^2|R_{ns}(0)|^2|R_{ms}(0)|^2\sqrt{s^2-2s(m_3^2+m_4^2)+(m_3^2-m_4^2)^2}
}{81m_c^2s^2}\int^1_{-1}|\bar{M}|^2 d\cos\theta,
\end{eqnarray}
where $\theta$ is the scattering angle between $\vec{p_1}$ and
$\vec{p_3}$, $|\bar{M}|^2$ is as follows
\begin{equation}
|\bar{M}_{\eta_c\psi}|^2=\frac{16384m_c^2(t^2+u^2-32m_c^4)}{(8m_c^2
- t - u)^5}.
\end{equation}
Here the Mandelstam variables are defined as
\begin{equation}
s=(p_1+p_2)^2,
\end{equation}
\begin{equation}
t=(p_3-p_1)^2=\frac{m_3^2+m_4^2-s}{2}+\frac{\sqrt{s^2-2s(m_3^2+m_4^2)+(m_3^2-m_4^2)^2}}{2}\cos\theta,
\end{equation}
\begin{equation}
u=(p_3-p_2)^2=\frac{m_3^2+m_4^2-s}{2}-\frac{\sqrt{s^2-2s(m_3^2+m_4^2)+(m_3^2-m_4^2)^2}}{2}\cos\theta,
\end{equation}

The cross sections for the double S-wave charmonium production are
listed as follows

\begin{equation}
\sigma(e^++e^-\rightarrow J/\psi(1S)+ \eta_{c}(1S)[\eta_{c}(2S),
\eta_{c}(3S)])=5.5[3.6, 3.1]~\rm{fb},
\end{equation}
\begin{equation}
\sigma(e^++e^-\rightarrow \psi(2S)+ \eta_{c}(1S)[\eta_{c}(2S),
\eta_{c}(3S)])=3.6[2.3, 2.0]~\rm{fb},
\end{equation}
where the values in the brackets are the cross sections respectively
for $\eta_c(2S)$ and $\eta_c(3S)$ with recoiling $\psi$ mesons.

\subsection{The $\psi(nS)\chi_{cJ}(mP) (J=0,1,2)$ production}

In the one spin-triplet S-wave (nS) and one spin-triplet P-wave (mP)
case, the cross section for $e^++e^-\rightarrow \gamma^{*}
\rightarrow H_1(nS)+H_2(mP)$ process reads
\begin{eqnarray}
\label{jskc} &&\sigma(e^+(p_1)+e^-(p_2)\rightarrow
H_1(p_3)+H_2(p_4))=\nonumber
\\ && \frac{2\pi\alpha^2\alpha_s^2|R_{ns}(0)|^2|R_{mp}'(0)|^2\sqrt{s^2-2s(m_3^2+m_4^2)+(m_3^2-m_4^2)^2}
}{27m_c^2s^2}\int^1_{-1}|\bar{M}|^2 d\cos\theta.
\end{eqnarray}
Here for the production of spin-triplet states $\psi \chi_{cJ}$,
$|\bar{M}|^2$ is given in Eq.~(\ref{kc0}) for $\chi_{c0}$,  in
 Eq.~(\ref{kc1})for $\chi_{c1}$ and Eq.~(\ref{kc2}) for
$\chi_{c2}$,
\begin{eqnarray}
\label{kc0}
|\bar{M}_{\psi\chi_{c0}}|^2&=&2048(90112m_c^{10}-74752m_c^8t-74752m_c^8u+23360m_c^6t^2
+43136m_c^6tu\nonumber\\&+&23360m_c^6u^2-3152m_c^4t^3-7600m_c^4t^2u-
7600m_c^4tu^2-3152m_c^4u^3\nonumber\\&+&162m_c^2t^4+444m_c^2t^3u+564m_c^2t^2u^2
+444m_c^2tu^3+162m_c^2u^4\nonumber\\&-&t^4u-3t^3u^2-3t^2u^3-tu^4)/(3
s^7m_c^2).
\end{eqnarray}
\begin{eqnarray}
\label{kc1} &&\hspace{-2.5cm} |\bar{M}_{\psi\chi_{c1}
}|^2=32768(1792m_c^8+256m_c^6t+256m_c^6u-56m_c^4t^2
-64m_c^4tu-56m_c^4u^2-4m_c^2t^3\nonumber\\ &&
-20m_c^2t^2u-20m_c^2tu^2-4m_c^2u^3+t^4+2t^3u+2t^2u^2+2tu^3+u^4)/s^7,
\end{eqnarray}
\begin{eqnarray}
\label{kc2} &&\hspace{-1.5cm} |\bar{M}_{\psi\chi_{c2}
}|^2=4096(145408m_c^{10}-1024m_c^8t-1024m_c^8u-2368m_c^6t^2
-6400m_c^6tu-2368m_c^6u^2\nonumber\\
&&+16m_c^4t^3-208m_c^4t^2u-208m_c^4tu^2
+16m_c^4u^3+24m_c^2t^4+72m_c^2t^3u+96m_c^2t^2u^2\nonumber\\
&&+72m_c^2tu^3 +24m_c^2u^4-t^4u-3t^3u^2-3t^2u^3-tu^4)/(3s^7m_c^2).
\end{eqnarray}

For the known spin-triplet P-wave states $\chi_{cJ}(1P)(J=0,1,2)$ we
find
\begin{equation}
\label{kc1p}
\sigma(e^++e^-\rightarrow J/\psi(1S)+
\chi_{c0}(1P)[\chi_{c1}(1P), \chi_{c2}(1P)])=6.7[1.1, 1.6]~\rm{fb},
\end{equation}
\begin{equation}
\label{kc1p} \sigma(e^++e^-\rightarrow \psi(2S)+
\chi_{c0}(1P)[\chi_{c1}(1P), \chi_{c2}(1P)])=4.4[0.74, 1.1]~\rm{fb},
\end{equation}

For the exited spin-triplet 2P states $\chi_{cJ}(2P)(J=0,1,2)$,
which are to be searched for, we find
\begin{equation}
\label{kc2p}
\sigma(e^++e^-\rightarrow J/\psi(1S)+
\chi_{c0}(2P)[\chi_{c1}(2P), \chi_{c2}(2P)])=9.1[1.6, 2.2]~\rm{fb},
\end{equation}
\begin{equation}
\sigma(e^++e^-\rightarrow \psi(2S)+ \chi_{c0}(2P)[\chi_{c1}(2P),
\chi_{c2}(2P)])=5.9[1.0, 1.4]~\rm{fb}.
\end{equation}

In Eqs. (19--22), we see that the cross sections for the excited
$2P$ states $\chi_{cJ}(2P)$ are somewhat larger than that for the
corresponding $1P$ states $\chi_{cJ}(1P)$ in the nonrelativistic
limit. This numerical result is due to the fact that we have chosen
the first derivative of the wave function at the origin for the $2P$
states to be larger than that for the $1P$ states, and actually the
former could be slightly smaller than the latter, depending on the
potentials that are used (see the values in the QCD (BT) model and
other models in Ref.\cite{wf}). Furthermore, another important
effect comes from the relativistic corrections, which may lower the
cross sections for the $\chi_{cJ}(2P)$ states. E.g., if we take the
charm quark mass to be $2m_c=M(2P)\approx 4~GeV$ for the 2P states,
then the cross sections will be substantially lower. Despite of
these uncertainties, we expect that the cross sections for the $2P$
states should be comparable to that for the $1P$ states.

\subsection{The  $\eta_c(nS) h_c(mP)$ production}

To search for the spin singlet P-wave charmonium $h_c$ is certainly
interesting. Recently, CLEO has found evidence for $h_c$ in the
$\psi(2S)\rightarrow\pi^0 h_c$ decay followed by $h_c\rightarrow
\gamma\eta_c$ with a mass of $M(h_c)=3524.4\pm 0.6\pm 0.4~MeV$ and
hyperfine splitting of about 1.0 MeV measured in both the $\eta_c$
exclusive and inclusive analysis\cite{cleo}. It will also be
interesting to search for the $h_c$ in  $e^+e^-$ annihilation at
$\sqrt{s}=10.6$ GeV in the recoil spectra of charge parity C=+1
states such as $\eta_c, \chi_{c0}, \chi_{c1}, \chi_{c2}$. For the
production of one spin-singlet S-wave state $\eta_c(nS)$ and one
spin-singlet P-wave state $h_c(mP)$, differing from \cite{bl}, we
find $|\bar{M}|^2$ to be not vanishing but given by
Eq.~(\ref{echc}):
\begin{eqnarray}
\label{echc} |\bar{M}_{\eta_c
h_c}|^2&=&2048(8192m_c^{10}+1024m_c^8t+1024m_c^8u+320m_c^6t^2+128m_c^6tu+320m_c^6u^2\nonumber\\&
-&16m_c^4t^3-112m_c^4t^2u-112m_c^4tu^2-16m_c^4u^3+2m_c^2t^4-4m_c^2t^3u-12m_c^2t^2u^2\nonumber\\&
-&4m_c^2tu^3+2m_c^2u^4-t^4u-3t^3u^2-3t^2u^3-tu^4)/((8m_c^2-t-u)^7m_c^2).
\end{eqnarray}

For the spin-singlet P-wave states $h_c(1P)$ and $h_c(2P)$ we find
\begin{equation}
\sigma(e^++e^-\rightarrow \eta_{c}(1S)+
h_{c}(1P)[h_{c}(2P)])=0.73[0.99]~\rm{fb},
\end{equation}
\begin{equation}
\sigma(e^++e^-\rightarrow \eta_{c}(2S)+
h_{c}(1P)[h_{c}(2P)])=0.48[0.65]~\rm{fb},
\end{equation}
\begin{equation}
\sigma(e^++e^-\rightarrow \eta_{c}(3S)+
h_{c}(1P)[h_{c}(2P)])=0.41[0.56]~\rm{fb}.
\end{equation}

\subsection{The $\chi_{cJ}(nP)(J=0,1,2)h_c(mP)$ production }

In the two P-wave (nP and mP) case, the cross section for
$e^++e^-\rightarrow \gamma^{*} \rightarrow H_1(nP)+H_2(mP)$ is
\begin{eqnarray}
\label{jsetc} &&\sigma(e^+(p_1)+e^-(p_2)\rightarrow
H_1(p_3)+H_2(p_4))=\nonumber\\
&&\frac{2\pi\alpha^2\alpha_s^2|R_{p}'(0)|^4\sqrt{s^2-2s(m_3^2+m_4^2)+(m_3^2-m_4^2)^2}
}{9m_c^2s^2}\int^1_{-1}|\bar{M}|^2 d\cos\theta.
\end{eqnarray}
For the $\chi_{cJ}h_c$ production $|\bar{M}|^2$ reads
\begin{eqnarray}
\label{kc0hc} |\bar{M}_{\chi_{c0}
h_c}|^2&=&16384(t^2+u^2-32m_c^4)(16m_c^2 - 3t - 3u)^2/(3(8m_c^2 - t
- u)^7m_c^2),
\end{eqnarray}
\begin{eqnarray}
\label{kc1hc} |\bar{M}_{\chi_{c1} h_c}|^2&=&- 4096(8192m_c^{10} -
3072m_c^8t - 3072m_c^8u + 2048m_c^6t^2 + 3584m_c^6tu \nonumber\\
&+&2048m_c^6u^2 - 16m_c^4t^3 + 144m_c^4t^2u + 144m_c^4tu^2 -
16m_c^4u^3 - 52m_c^2t^4 \nonumber\\
&-& 128 m_c^2t^3u - 152m_c^2t^2u^2 - 128m_c^2tu^3 - 52m_c^2u^4 +
t^4u \nonumber\\&+& 3t^3u^2 + 3t^2u^3 + tu^4) /((8m_c^2 - t -
u)^7m_c^4),
\end{eqnarray}
\begin{eqnarray}
\label{kc2hc} |\bar{M}_{\chi_{c2} h_c}|^2&=& - 8192(544m_c^4 +
72m_c^2t + 72m_c^2u + 3t^2 + 6tu + 3u^2)\nonumber\\&\times&(32m_c^4
- t^2 - u^2)/(3(8m_c^2 - t - u)^7m_c^2),
\end{eqnarray}
and the corresponding cross sections are
\begin{equation}
\sigma(e^++e^-\rightarrow \chi_{c0}(1P)+
h_{c}(1P)[h_{c}(2P)])=0.22[0.31]~\rm{fb},
\end{equation}
\begin{equation}
\sigma(e^++e^-\rightarrow \chi_{c1}(1P)+
h_{c}(1P)[h_{c}(2P)])=1.0[1.4]~\rm{fb},
\end{equation}
\begin{equation}
\sigma(e^++e^-\rightarrow \chi_{c2}(1P)+
h_{c}(1P)[h_{c}(2P)])=0.063[0.085]~\rm{fb}.
\end{equation}

\subsection{The $\psi(nS)^1D_2(mP)$ production}

 The cross section for
${e^++e^-\rightarrow \gamma^{*} \rightarrow H_1(nS)+H_2(mD)}$
process is formulated as
\begin{eqnarray}
\label{sss} &&\sigma(e^+(p_1)+e^-(p_2)\rightarrow
H_1(p_3)+H_2(p_4))=\nonumber\\
&&\frac{2\pi\alpha^2\alpha_s^2|R_{S}(0)|^2|R^{\prime\prime}_{D}(0)|^2\sqrt{s^2-2s(m_3^2+m_4^2)+(m_3^2-m_4^2)^2}
}{27m_c^2s^2}\int^1_{-1}|\bar{M}|^2 d\cos\theta,
\end{eqnarray}

\begin{eqnarray}
\label{j1d2} \mid \bar{M}_{J/\psi ^1D_2} \mid &=& - 327680(16 m_c^2
- s)^2 (512 m_c^6 - 32 m_c^4 s - 128 m_c^4 t - 128 m_c^4 u\nonumber\\
&+& 8 m_c^2 s t + 8 m_c^2 s u + 8 m_c^2 t^2 + 16 m_c^2 t u + 8 m_c^2
u^2 - s t^2 - s u^2)/(3 m_c^2 s^8),
\end{eqnarray}
and the numerical result is
\begin{equation}
\sigma(e^++e^-\rightarrow J/\psi(1S)[2s]+ ^1D_2)=0.19[0.12]~\rm{fb}.
\end{equation}

\subsection{The $\chi_{cJ}(nP)(J=0,1,2)^3D_J'(mD)(J'=1,2,3)$ production }

The cross section for ${e^++e^-\rightarrow \gamma^{*} \rightarrow
H_1(nP)+H_2(mD)}$ process is formulated as
\begin{eqnarray}
\label{sss} &&\sigma(e^+(p_1)+e^-(p_2)\rightarrow
H_1(p_3)+H_2(p_4))=\nonumber\\
&&\frac{2\pi\alpha^2\alpha_s^2|R_{P}(0)|^2|R^{\prime\prime}_{D}(0)|^2\sqrt{s^2-2s(m_3^2+m_4^2)+(m_3^2-m_4^2)^2}
}{27m_c^2s^2}\int^1_{-1}|\bar{M}|^2 d\cos\theta,
\end{eqnarray}
\begin{eqnarray}
\label{k13d1} \mid \bar{M}_{\chi_{c1}~ ^3D_1} \mid &=& 20480
(113393664 m_c^{12} - 49483776 m_c^{10} s - 28348416 m_c^{10} t -
28348416 m_c^{10} u \nonumber\\&+& 7436288 m_c^8 s^2 + 12601344
m_c^8 s t + 12601344 m_c^8 s u + 1829376 m_c^8 t^2 \nonumber\\&+&
3428352 m_c^8 t u + 1829376 m_c^8 u^2 - 413120 m_c^6 s^3 - 912960
m_c^ 6 s t^2\nonumber\\&-& 1932032 m_c^6 s^2 t - 1932032 m_c^6 s^2 u
- 1324416 m_c^6 s t u - 912960 m_c^6 s u^2 \nonumber\\&+& 5600 m_c^4
s^4 + 114816 m_c^4 s^3 t + 114816 m_c^4 s^3  u + 165024 m_c^4 s^2
t^2 + 152960 m_c^4 s^2 t u \nonumber\\&+& 165024 m_c^4 s^2 u^2 + 144
m_c^2 s^5 - 2312 m_c^2 s^4 t - 2312 m_c^2 s^4 u - 12184 m_c^2 s^3
t^2 \nonumber\\&-& 4336 m_c ^2 s^3 t u - 12184 m_c^2 s^3 u^2 + 289
s^4 t^2 + 289 s^4 u^2)/(5 m_c^4 s^9),
\end{eqnarray}
\begin{eqnarray}
\label{k13d2} \mid \bar{M}_{\chi_{c1}~ ^3D_2} \mid &=&2560 (58589184
m_c^{14} - 2293760 m_c^{12} s - 14647296 m_c^{12} t- 4286464
m_c^{10} s^2 \nonumber\\&-& 14647296 m_c^{12} u  + 6922240 m_c^{10}
s t + 6922240 m_c^{10} s u + 2502656 m_c^{10} t^2 \nonumber\\&-&
1343488 m_c ^{10} t u
+ 2502656 m_c^{10} u^2 + 576512 m_c^8 s^3 - 1119744 m_c^8 s^2 t\nonumber\\
& -&  938496 m_c^8  s t^2  - 1119744 m_c^8 s^2 u+146432 m_c^8 s t u
- 938496 m_c^8 s u^2  \nonumber\\&+& 4096 m_c^6 s^4 + 72192 m_c^6
s^3 t - 1248 m_c^4 s^5- 5984 m_c^4 s^3 t^2 - 6080 m_c^4 s^3 t
u\nonumber\\&+& 72192 m_c^6 s^3 u + 224 m_c^2 s ^4 t^2  + 119488
m_c^6 s^2 t^2+ 40960 m_c^6 s^2 t u \nonumber\\&-& 1792 m_c^4 s^4 t+
119488 m_c ^6 s^2 u^2  - 1792 m_c^4 s^4 u  - 5984 m_c^4 s^3 u^2
\nonumber\\&- &48 m_c^2 s^6  + 224 m_c^2 s^4 u^2 + 3 s^7 - 3 s^5 t^2
+ 6 s^5 t u - 3 s^5 u^2)/(m_c^6 s^9),
\end{eqnarray}
\begin{eqnarray}
\label{k13d3} \mid \bar{M}_{\chi_{c1}~ ^3D_3} \mid &=&40960 (7446528
m_c^{10} - 756224 m_c^8 s - 1861632 m_c^8 t + 393856 m_c^6 s t
\nonumber\\&-& 1861632 m_c^8 u - 62208 m_c^6 s^2 + 393856 m_c^6 s u
+ 167552 m_c^6 t^2 + 130304 m_c^6 t u \nonumber\\&+& 167552 m_c^6
u^2 + 9632 m_c^4 s^3 - 27968 m_c^4 s^2 t - 27968 m_c^4 s^2 u - 39568
m_c^4 s t^2 \nonumber\\&-& 19328 m_c^4 s t u - 39568 m_c^4 s u^2 -
288 m_c^2 s^4 + 664 m_c^2 s ^3 t + 664 m_c^2 s^3 u \nonumber\\& +&
752 m_c ^2 s^2 t u + 3120 m_c^2 s^2 t^2+ 3120 m_c^2 s^2 u^2 - 83 s^3
t^2 \nonumber\\&-& 83 s^3 u^2) (16 m_c^2 - s)/(5m_c^4 s^9),
\end{eqnarray}
\begin{eqnarray}
\label{k23d1} \mid \bar{M}_{\chi_{c2}~ ^3D_1} \mid &=&640
(29807345664 m_c^{14} - 6599344128 m_c^{12} s - 7451836416 m_c^ {12}
t\nonumber\\& -& 7451836416 m_c^{12} u + 182222848 m_c^{10} s^2 +
2217541632 m_c^{10} s t + 2217541632 m_c^{10} s u \nonumber\\&+&
647626752 m_c^{10} t u + 607666176  m_c^{10} t^2 + 607666176
m_c^{10} u^2 + 43309056 m_c^8 s^3 \nonumber\\&-& 233439232 m_c^8 s^2
t - 233439232 m_c^8 s^2 u - 205805568 m_c^8 s t^2 - 142774272 m_c^8
s t u \nonumber\\&-& 205805568 m_c^8 s u^2 - 3057664 m_c^6 s^4 +
10221568 m_c^6 s^3 t + 10221568 m_c^6 s^3 u \nonumber\\&+& 24313856
m_c ^6 s^2 t^2 + 9732096 m_c^6 s^2 t u + 24313856 m_c^6 s ^2 u^2 +
66368 m_c^4 s^5 \nonumber\\&-& 163072 m_c^4 s^4 t - 163072
 m_c^4 s^4 u - 1174848 m_c^4 s^3 t^2 - 400 m_c^2 s^6 \nonumber\\&-& 205696 m_c^4 s^3  t u -
1174848 m_c^4 s^3 u^2 + 20768 m_c ^2 s^4 t^2 \nonumber\\&-& 768
m_c^2 s^4 t u + 20768 m_c^2 s^4 u^2
 + s^7 - s^5 t^2 + 2 s^5 t u - s^5 u^2)/(15 m_c^6 s^9),
\end{eqnarray}
\begin{eqnarray}
\label{k3d2} \mid \bar{M}_{\chi_{c2}~ ^3D_2} \mid &=&81920(13979648
m_c^{12} - 1540096 m_c^{10} s - 3494912 m_c^{10} t - 3494912
m_c^{10} u \nonumber\\&-& 62400 m_c^8 s^2 + 724224 m_c^8 s t +
724224 m_c^8 s u + 303232 m_c^8 t^2 + 267264 m_c^8 t u
\nonumber\\&+& 303232 m_c^8 u^2 + 14208 m_c^6 s^3 - 55440 m_c^6 s^2
t - 55440 m_c^6 s^2 u - 71376 m_c^6 s t^2 \nonumber\\&-& 38304 m_c^6
s t u - 71376 m_c^6 s u^2 - 624 m_c^4 s^4 + 1872 m_c^4 s^3 t + 1872
m_c^4 s^3 u \nonumber\\&+& 6018 m_c^4 s^2 t^2 + 1824 m_c^4 s^2 t u +
6018 m_c^4 s^2 u^2 + 9 m_c^2 s^5
 - 24 m_c^2 s^4 t \nonumber\\&-& 24 m_c^2 s^4 u - 219 m_c^2 s^3 t^
2 - 30 m_c^2 s^3 t u - 219 m_c^2 s^3 u^2 \nonumber\\&+& 3 s^4 t^ 2 +
3 s^4 u^2)/( m_c^4 s^9),
\end{eqnarray}
\begin{eqnarray}
\label{k3d2} \mid \bar{M}_{\chi_{c2}~ ^3D_3} \mid &=&5120 (883687424
m_c^{14} - 77856768 m_c^{12} s - 220921856 m_c^{12} t \nonumber\\&-&
220921856 m_c^{12} u - 2650112 m_c^{10} s^2 + 40353792 m_c^ {10} s t
+ 40353792 m_c^{10} s u \nonumber\\&+& 19030016 m_c^{10} t^2 +
17170432 m_c^{10} t u + 19030016 m_c^{10} u^2 + 1043456 m_c^8 s^3
\nonumber\\&-& 2798592 m_c^8 s^2 t - 2798592 m_c^8 s^2 u - 3923968
m_c^8 s t^2 - 2240512 m_c^8 s t u \nonumber\\&-& 3923968 m_c^8 s u^2
- 95744 m_c^6 s^4 + 89088 m_c^6 s^3 t + 89088 m_c ^6 s^3 u
\nonumber\\&+& 332416 m_c^6 s^2 t^2 + 34816 m_c^6 s^2 t u + 332416
m_c^6 s^2 u^2 + 3968 m_c^4 s^5 \nonumber\\&-& 1152 m_c^4 s^4 t -
1152 m_c^4 s^4 u - 12928 m_c^4 s^3 t^2 + 3584 m_c^4 s^3 t u
\nonumber\\&-& 12928 m_c^4 s^3 u^2 - 80 m_c^2 s^6 + 208 m_c^2 s^4
t^2 - 128 m_c^2 s^4 t u + 208 m_c^2 s^4 u^2 \nonumber\\&+& s^7 - s^5
t^2 + 2 s^5 t u - s^5 u^2)/(5 m_c^6 s^9).
\end{eqnarray}

The numerical results are listed as follows, where $\delta_1\equiv
{^3D_1},~\delta_2\equiv {^3D_2},~\delta_3\equiv {^3D_3}$,
\begin{equation}
\sigma(e^++e^-\rightarrow \chi_{c1}[\chi_{c2}]+
\delta_1)=0.080[0.041]~\rm{fb},
\end{equation}
\begin{equation}
\sigma(e^++e^-\rightarrow \chi_{c1}[\chi_{c2}]+
\delta_2)=0.099[0.084]~\rm{fb},
\end{equation}
\begin{equation}
\sigma(e^++e^-\rightarrow \chi_{c1}[\chi_{c2}]+
\delta_3)=0.041[0.0099]~\rm{fb},
\end{equation}

As a summary of the above results, we show the differential cross
sections (the angular distribution functions) for different double
charmonium production processes at leading order in NRQCD (QED
contributions are not included) in Table I, and the corresponding
graphs in Fig.~1-7.

\section{QED processes including  $J^{PC}=1^{--}$ states}
  Since the $J^{PC}=1^{--}~c\bar{c}$ can be produced via a single photon,
the QED contribution may be significant or even comparable to the
QCD contribution in some exclusive processes involving one
$J^{PC\mathfrak{S}}=1^{--}$ charmonium state. These QED effects are
considered in \cite{bl,bler}. There are six Feynman diagrams for the
QED process at $\alpha^{4}$ order, and only two represent
$\gamma^*\rightarrow
c\bar{c}\gamma^*\rightarrow(c\bar{c})_{1^{--}}c\bar{c}$, which is
dominant and has been calculated in \cite{bl,bler}. In this paper we
include all the eight diagrams to get the full result at order
$\alpha^{4}$, though the contributions of other six diagrams are
numerically small. Using the notation in section II, we re-express
below the analytical formulas of the exclusive processes including
both QCD and QED contributions.

\subsection{The $\psi(nS)\eta_c(mS)$ production}

  The cross section for $e^++e^-\rightarrow \gamma^{*} \rightarrow
H_1(nS)+H_2(mS)$ is now changed to

\begin{eqnarray}
\label{jsetcE} &&\sigma(e^+(p_1)+e^-(p_2)\rightarrow
H_1(p_3)+H_2(p_4))=\nonumber\\
&&\frac{\alpha^2|R_{ns}(0)|^2|R_{ms}(0)|^2\sqrt{s^2-2s(m_3^2+m_4^2)+(m_3^2-m_4^2)^2}
}{288m_c^2\pi s^2}\int^1_{-1}|\bar{M}|^2 dx,
\end{eqnarray}
where $x=\cos \theta$ and $\theta$ is the scattering angle between
$\overrightarrow{p_{1}}$ and $ \overrightarrow{p_{3}}$. And
$|\bar{M}|^{2}$is
\begin{equation}
|\bar{M}|^{2}=\frac{2048(s-16m_c^2)(16\alpha
m_c^2+48\alpha_{s}mc^2+3\alpha s)^2}{81m_c^2 s^4}
\end{equation}
  The numerical results become
\begin{equation}
\sigma(e^++e^-\rightarrow J/\psi(1S)+ \eta_{c}(1S)[\eta_{c}(2S),
\eta_{c}(3S)])=6.6[4.3, 3.7]~\rm{fb},
\end{equation}
\begin{equation}
\sigma(e^++e^-\rightarrow \psi(2S)+ \eta_{c}(1S)[\eta_{c}(2S),
\eta_{c}(3S)])=4.3[2.8, 2.4]~\rm{fb},
\end{equation}
where the values in the brackets are the cross sections respectively
for $\eta_c(2S)$ and $\eta_c(3S)$ with recoiled $\psi$ mesons.

\subsection{The $\psi(ns)+\chi_{cJ}(mp)$ Production}

The cross section for $e^++e^-\rightarrow \gamma^{*} \rightarrow
H_1(nS)+H_2(mS)$ is changed to
\begin{eqnarray}
\label{jskcE} &&\sigma(e^+(p_1)+e^-(p_2)\rightarrow
H_1(p_3)+H_2(p_4))=\nonumber\\
&&\frac{\alpha^2|R_{ns}(0)|^2|R_{mp}'(0)|^2\sqrt{s^2-2s(m_3^2+m_4^2)+(m_3^2-m_4^2)^2}
}{96m_c^2\pi s^2}\int^1_{-1}|\bar{M}|^2 dx.
\end{eqnarray}
And now  $|\bar{M}|^2$ for $\psi \chi_{c0}, \psi \chi_{c1}, \psi
\chi_{c2}$ production, are given in Eq.~(\ref{kc0E}),
 Eq.~(\ref{kc1E}), Eq.~(\ref{kc2E}) respectively.
\begin{eqnarray}
\label{kc0E}
|\bar{M}_{\psi\chi_{c0}}|^2&&=(2048\pi^2((589824(x^2-1)m_c^{10}
-4096s(73x^2+25)m_c^8-2048s^2(x^2-1)m_c^6\nonumber\\&&+128s^3(25x^2+17)m_c^4
+16s^4(x^2-1)m_c^2-9s^5(x^2+1))\alpha^2\nonumber\\&&+48\alpha_s
m_c^2(73728(x^2-1)m_c^8-3584s(13x^2+1)m_c^6+64s^2(71x^2+55)m_c^4\nonumber\\&&
+56s^3(5x^2+1)m_c^2- s^4(25x^2+29))\alpha+144\alpha_s^2
m_c^2(36864(x^2-1)m_c^8\nonumber\\&&
-256s(109x^2-11)m_c^6+128s^2(41x^2+22)m_c^4-4s^3(61x^2+101)m_c^2\nonumber\\&&+s^4(x^2-1))))/(243m_c^4
s^6).
\end{eqnarray}
\begin{eqnarray}
\label{kc1E} |\bar{M}_{\psi\chi_{c1}
}|^2&&=(4096\pi(192\alpha\alpha_s(9216(x^2-1)m_c^6-64s(31x^2+19)m_c^4+8s^2(9x^2-19)m_c^2
\nonumber\\&&+s^3(x^2+25))m_c^4+2304\alpha_s^2(1152(x^2-1)m_c^6-8s(49x^2+1)m_c^4
\nonumber\\&&+4s^2(9x^2+5)m_c^2-s^3(x^2+1))m_c^4+\alpha^2(294912(x^2-1)m_c^{10}
\nonumber\\&&-2048s(13x^2+37)m_c^8-19456s^2
m_c^6+32s^3(x^2-23)m_C^4+144s^4 x^2
m_c^2\nonumber\\&&-9s^5(x^2+1))))/(81m_c^4 s^6).
\end{eqnarray}
\begin{eqnarray}
\label{kc2E} |\bar{M}_{\psi\chi_{c2}
}|^2&&=(4096\pi^2((1474560(x^2-1)m_c^{10}+2048s(31x^2-233)m_c^8-1024s^2(32x^2+1)m_c^6
\nonumber\\&&-32s^3(35x^2-101)m_c^4+16s^4(19x^2-10)-9s^5(x^2+1))\alpha^2\nonumber\\&&+48\alpha_s
m_c^2(184320(x^2-1)m_c^8-256s(59x^2+143)m_c^6-32s^2(59x^2-89)m_c^4\nonumber\\&&+4s^3(61x^2+29)m_c^2
-s^4(7x^2+11))\alpha+144\alpha_s^2
m_c^2(92160(x^2-1)m_c^8\nonumber\\&&-128s(149x^2+53)m_c^6+64s^2(25x^2+23)m_c^4
-32s^3(2x^2+1)m_c^2\nonumber\\&&+s^4(x^2-1))))/(243m_c^4 s^6).
\end{eqnarray}

The numerical results of spin-triplet P-wave states
$\chi_{cJ}(1P)(J=0,1,2)$ are
\begin{equation}
\label{kc1pE} \sigma(e^++e^-\rightarrow J/\psi(1S)+
\chi_{c0}(1P)[\chi_{c1}(1P), \chi_{c2}(1P)])=6.9[1.0, 1.8]~\rm{fb},
\end{equation}
\begin{equation}
\sigma(e^++e^-\rightarrow \psi(2S)+ \chi_{c0}(1P)[\chi_{c1}(1P),
\chi_{c2}(1P)])=4.5[0.7, 1.1]~\rm{fb},
\end{equation}

And for the exited spin-triplet 2P states $\chi_{cJ}(2P)(J=0,1,2)$,
the results turn to be
\begin{equation}
\label{kc2pE} \sigma(e^++e^-\rightarrow J/\psi(1S)+
\chi_{c0}(2P)[\chi_{c1}(2P), \chi_{c2}(2P)])=9.4[1.4, 2.4]~\rm{fb},
\end{equation}
\begin{equation}
\sigma(e^++e^-\rightarrow \psi(2S)+ \chi_{c0}(2P)[\chi_{c1}(2P),
\chi_{c2}(2P)])=6.2[0.9, 1.6]~\rm{fb}.
\end{equation}

\subsection{The $\psi(ns)+1D2$ Production}
The cross section for ${e^++e^-\rightarrow \gamma^{*} \rightarrow
H_1(nS)+H_2(mD)}$ process is
\begin{eqnarray}
\label{sssE} &&\sigma(e^+(p_1)+e^-(p_2)\rightarrow
H_1(p_3)+H_2(p_4))=\nonumber\\
&&\frac{5\alpha^2|R_{S}(0)|^2|R^{\prime\prime}_{D}(0)|^2\sqrt{s^2-2s(m_3^2+m_4^2)+(m_3^2-m_4^2)^2}
}{192 m_c^2 \pi s^2}\int^1_{-1}|\bar{M}|^2 dx,
\end{eqnarray}
and
\begin{eqnarray}
\label{j1d2E} {\mid \bar{M}_{J/\psi ^1D_2} \mid}^2
&=&\frac{4096\pi^2(s-16m_c^2)^3(32\alpha m_c^2+96\alpha_s m_c^2+3s
\alpha )^2(x^2+1)}{243m_c^6 s^6},
\end{eqnarray}
and the numerical result becomes
\begin{equation}
\sigma(e^++e^-\rightarrow J/\psi(1S)[2s]+ ^1D_2)=0.21[0.13]~\rm{fb}.
\end{equation}

\subsection{The $\eta_c(ns)+\psi_{3D1}$ Production}

The cross section for ${e^++e^-\rightarrow \gamma^{*} \rightarrow
H_1(nS)+H_2(mD)}$ process is formulated as
\begin{eqnarray}
\label{ssDE} &&\sigma(e^+(p_1)+e^-(p_2)\rightarrow
H_1(p_3)+H_2(p_4))=\nonumber\\
&&\frac{5\alpha^2|R_{S}(0)|^2|R^{\prime\prime}_{D}(0)|^2\sqrt{s^2-2s(m_3^2+m_4^2)+(m_3^2-m_4^2)^2}
}{192 m_c^2 \pi s^2}\int^1_{-1}|\bar{M}|^2 dx,
\end{eqnarray}
and
\begin{eqnarray}
\label{e3d1E} |\bar{M}_{\eta_c ~^3D_1}|^2
&=&512\pi^2(s-16m_c^2)(48\alpha_s (64m_c^2-9s)m_c^2+\alpha
(1024m_c^4-144sm_c^2-15s^2 ))^2\nonumber\\&&(x^2+1)/(1215m_c^6 s^6)
,
\end{eqnarray}
and the numerical result becomes
\begin{equation}
\sigma(e^++e^-\rightarrow \eta_c(1S)+ ^3D_1)=0.17~\rm{fb}.
\end{equation}

\subsection{conclusion}

  In this section we have considered the QED contribution to the exclusive
processes involving one $J^{pc}=1^{--}$ charmonium state (e.g.,
$\psi(nS)$ and $\psi_{3D1}$). We find that in general by adding QED
contribution the cross section can be changed by the order of ten
percent. We list the results below with $\sqrt{s}=10.6$GeV,
$\alpha_{s}=0.26, m_c=1.5GeV$. The cross sections of
$\psi(ns)+\eta_c(ms)$ are increased by 20 percent, that of
$\psi(ns)+\chi_{cJ}$ are increased by 4,~-5,~8 percent for $J=0,1,2$
respectively, that of $\psi(ns)+^1D_2$ is increased by $11$ percent,
and that of $\eta_{c}(ns)+ ^3D_1$ is increased by $9$ percent. In
\cite{bl} the authors also considered the QED process but only the
two dominant diagrams were included. Our results are in agreement
with their results in most processes when we choose the same
parameters as theirs, except for the process of
$\eta_{c}+\psi(^3D_1)$ production, for which they obtained 41\%
enhancement with QED effects whereas we get 19\% with the same
parameters. Here our analytical expression also differs from theirs.

We show the differential cross sections (the angular distribution
functions) for different double charmonium production processes
including both QCD and QED contributions in Table II, and the
corresponding graphs in Fig.~8. Ratios of production cross sections
of various double charmonia to that of $J/\psi+\eta_c$ in $e^+ e^-$
annihilation at $\sqrt{s}=10.6$~GeV are listed in Table III.

\section{Discussion and Summary}
In this paper, we make predictions for various double charmonia
production processes in $e^+e^-$ annihilation  at $\sqrt{s}=10.6$
GeV with $B$ factories, based on a complete leading order
calculation including both QCD and QED contributions. In particular,
we aim at searching for excited charmonium states in these
processes. From the obtained results we make the following
observations:

The calculated relative production rates for $e^+e^-\rightarrow
\psi(2S)\eta_c$, $\psi(2S)\chi_{c0}, \psi(2S)\eta_c(2S)$ are roughly
compatible with the new Belle measurements \cite{belle2} (see also
\cite{bl}), assuming the decay branching ratios into charged tracks
are comparable for $\eta_c, \eta_c(2S),$ and $ \chi_{c0}$.

The calculated relative production rates for $e^+e^-\rightarrow
J/\psi\chi_{c0}(2P), J/\psi\eta_c(3S)$ and $e^+e^-\rightarrow
\psi(2S)\chi_{c0}(2P), \psi(2S)\eta_c(3S)$ are large, and these two
states may be observable in the mass range $m_{2P}=3.90-4.00$~GeV
and $m_{3S}=3.95-4.10$~GeV respectively (see, e.g. \cite{godfrey}).
Both of them are above the OZI (Okubo-Zweig-Iizuka) allowed
thresholds, but unlike the $\chi_{c0}(2P)$, the $\eta_c(3S)$ can not
decay to $D\bar D$ pair, which may distinguish between these two
states experimentally.

The calculated relative production rate for $e^+e^-\rightarrow
h_c\eta_c$ is not zero. This differs from the result given in
\cite{bl}, but agrees with \cite{bler}. Hence the $h_c$ might be
observable via this channel with high statistics in the future.
Aside from this process, $e^+e^-\rightarrow  h_c\chi_{c1}$ is also
very hopeful in finding the $h_c$ meson, since the $\chi_{c1}$ has a
large branching ratio (larger than 30\%) decaying into $\gamma
J/\psi$, and the $J/\psi$ can be easily detected by the $\mu^+\mu^-$
signal. Moreover, the calculated cross section for
$e^+e^-\rightarrow h_c\chi_{c1}$ is about 1.0~fb, not very small and
much larger than that for $h_c\chi_{c0}$ and $h_c\chi_{c2}$
production rates.

We show the differential cross sections (the angular distribution
functions) for different double charmonium production processes in
Table I (not including QED contribution) and Table II (including
both QCD and QED contributions), and the corresponding graphs in
Fig.~1-7, and Fig.~8.

As a whole, our results agree with those in \cite{bl} and
\cite{bler} (also agree with our previous result for
$J/\psi\eta_c(\chi_{c0},\chi_{c1},\chi_{c2})$\cite{liu}), if using
the same parameters. For the QED part, in \cite{bl,bler} only two
dominant diagrams are taken into account, while in this paper all
diagrams are considered, but the numerical contributions of the
remaining four diagrams are small. However, there still exists a
difference in the result for the $e^+e^-\to\eta_{c}+\psi(^3D_1)$
production. As for numerical results, since we use a larger value of
the strong coupling constant $\alpha_s=0.26$ (corresponding to
$\mu=2m_c=3.0~GeV$ and $ \Lambda^{(4)}_{\overline{MS}}$=338~MeV)
than that in \cite{bl,bler}, our predicted cross sections are in
general larger than that in \cite{bl,bler}. Moreover, we use the
charmonium wave functions and their derivatives at the origin
(including the ground state and excited states) from the BT
potential model calculation\cite{wf} but not from the experimental
values of leptonic decay widths, etc., as in \cite{bl,bler}, and
this may further enlarge our predicted cross sections. These
parameters may not be the best choice, but, at present, since we do
not have enough available data for higher excited charmonium states,
using the potential model calculation may still be a reasonable and
tentative choice. We view these as theoretical uncertainties in our
approach for the leading order calculations.

As emphasized above, this paper aims at searching for excited
charmonium states in $e^+e^-$ annihilation  at $\sqrt{s}=10.6$ GeV
with $B$ factories. After the main part of the results were
presented in \cite{0408141}
(with a more complete leading order calculation including both QCD
and QED contributions for various double charmonium production
processes being added in its present form), a number of new
experimental and theoretical results have appeared recently.

1. The double charmonium production in $e^+e^-$ annihilation at B
factories has been confirmed by the BaBar Collaboration (see
Ref.\cite{babar}) with comparable cross sections to that observed by
Belle. Theoretically, the large gap between experiment and theory
could be largely narrowed by the next to leading order QCD radiative
corrections~\cite{zhang} and relativistic corrections (see,
Refs.\cite{bodwin,he} and references therein) in the framework of
nonrelativistic QCD, and other possible approaches (see,
e.g.\cite{ji}).

2. Belle has observed the X(3940), a new charmonium state or
charmonium-like state, in $e^+e^-\rightarrow J/\psi
\mathrm{X}(3940)$ with $M_\mathrm{X}=(3.940 \pm
0.012)~GeV$\cite{pakhlov}. While its main decay mode is
$\mathrm{X}(3940)\to D\overline {D^*}$, no signal is found for
$\mathrm{X}(3940)\to D\overline {D}$. This rules out the possibility
of X(3940) being the $\chi_{c0}(2P)$ state. Furthermore, the X(3940)
is unlikely to be the $\chi_{c1}(2P)$ state, since no signal for the
$\chi_{c1}(1P)$ is found. This is in line with our calculation,
which shows that the cross section for $e^+e^-\rightarrow
J/\psi\chi_{c1}(1P)$ is much smaller than $e^+e^-\rightarrow
J/\psi\eta_c$. Finally, the X(3940) could be the $\eta_c(3S)$ state.
According to our calculation (see Eq.(13)), the production cross
sections for $\eta_c(1S),~\eta_c(2S),\eta_c(3S)$ are respectively
5.5,~3.6~,3.1~fb, and the relative rates are roughly consistent with
the signal yields $N=235 \pm 26,~164 \pm 30,~149 \pm 33$ events for
$\eta_c(1S),~\eta_c(2S),~\mathrm{X}(3940)$\cite{pakhlov}. This might
be viewed as a support to interpreting the X(3940) as the
$\eta_c(3S)$. However, the remaining problem is how to understand
its low mass of X(3940) if it is the $\eta_c(3S)$, which is lower
than potential model predictions by 50-120~MeV. But this could be
explained by the coupled channel effects that the coupling of
$\eta_c(3S)$ to the $0^+$ and $0^-$ charmed meson pair (in S-wave)
will lower the mass of $\eta_c(3S)$\cite{ELQ}.

3. Belle has not observed the X(3872) in $e^+e^-\rightarrow J/\psi
\mathrm{X}(3872)$. This implies that the X(3872) is unlikely to be a
conventional $0^{-+}$ or $0^{++}$ charmonium, since in our
calculation they may have relatively large rates to be observed in
double charmonium production. On the other hand, a $1^{++}$ or
$2^{-+}$ charmonium could be possible for X(3872), since they have
relatively small production rates. Of course, the nature of X(3872)
needs clarifying by other more relevant experiments (see, e.g.
\cite{swanson06}), aside from the $e^+e^-$ annihilation processes.

4. Belle has very recently observed a new state, the X(4160), in the
process of double charm production $e^+e^-\to
J/\psi+\mathrm{X}(4160)$ followed by $\mathrm{X}(4160)\to
D^*\bar{D^*}$\cite{belle0707}. Possible interpretations for the
\rm{X}(4160) are discussed in\cite{chao07} with emphasized possible
assignments of $2^1D_2, \eta_c(4S), \chi_{c0}(3P)$ charmonium states
and related problems.

In conclusion, we find that the double charmonium production
processes in $e^+e^-$ annihilation  at $B$ factories are very useful
tools in searching for excited charmonium states or charmonium-like
states. The leading order NRQCD calculation might hopefully provide
a useful guide for the relative production rates, but not the
absolute rates themselves. A systematical study for the QCD
radiative corrections and relativistic corrections for different
processes are apparently needed. On the other hand, studies of
charmonium spectroscopy including  charmonium masses, decays, and
coupled channel effects, and the new type charmonium-like states are
also very desirable.



\begin{acknowledgments}
We thank P. Pakhlov for discussions on the Belle data and possible
implications, and E. Braaten and J. Lee for communications on some
calculations. This work was supported in part by the National
Natural Science Foundation of China (No. 10421503, No. 10675003),
the Key Grant Project of Chinese Ministry of Education (No. 305001),
and the Research Found for Doctorial Program of Higher Education of
China.
\end{acknowledgments}

\newpage
\begin{center}
\begin{table*}
\caption{Differential cross sections--angular distribution functions
for double charmonium production in $e^+ e^-$ annihilation at
$\sqrt{s}=10.6$~GeV (QED contributions not included; see text for
the input parameters).}
\begin{tabular}{|l|l|}
  \hline
  Differential cross section & Angular distribution function (fb)\\\hline
  $d\sigma(e^++e^-\rightarrow J/\psi +
\eta_{c}(1S))/dcos\theta$ & $2.06(1+cos^2\theta)$ \\\hline
  $d\sigma(e^++e^-\rightarrow J/\psi +
\eta_{c}(2S))/dcos\theta$ & $1.34(1+cos^2\theta)$ \\\hline
  $d\sigma(e^++e^-\rightarrow J/\psi + \eta_{c}(3S))/dcos\theta$ &
$1.16(1+cos^2\theta)$ \\\hline
  $d\sigma(e^++e^-\rightarrow J/\psi +
\chi_{c0})/dcos\theta$ & $3.09(1+0.252cos^2\theta)$ \\\hline
  $d\sigma(e^++e^-\rightarrow J/\psi +
\chi_{c1})/dcos\theta$ & $0.457(1+0.698cos^2\theta)$ \\\hline
  $d\sigma(e^++e^-\rightarrow J/\psi +
\chi_{c2})/dcos\theta$ & $0.870(1-0.198cos^2\theta)$ \\\hline
  $d\sigma(e^++e^-\rightarrow \eta_{c}+
hc(1p))/dcos\theta$ & $0.480(1-0.726cos^2\theta)$ \\\hline
  $d\sigma(e^++e^-\rightarrow \eta_{c}+
hc(2p))/dcos\theta$ & $0.653(1-0.726cos^2\theta)$ \\\hline
  $d\sigma(e^++e^-\rightarrow \chi_{c1}+
hc(1p))/dcos\theta$ & $0.483(1+0.190cos^2\theta)$ \\\hline
  $d\sigma(e^++e^-\rightarrow \chi_{c1}+
hc(2p))/dcos\theta$ & $0.657(1+0.190cos^2\theta)$ \\\hline
  $d\sigma(e^++e^-\rightarrow \chi_{c2}+
hc(1p))/dcos\theta$ & $0.0235(1+cos^2\theta)$ \\\hline
  $d\sigma(e^++e^-\rightarrow \chi_{c2}+
hc(2p))/dcos\theta$ & $0.0306(1+cos^2\theta)$ \\\hline
  $\sigma(e^++e^-\rightarrow J/\psi(1S)+
^1D_2)/dcos\theta$ & $0.0694(1+cos^2\theta)$ \\\hline
  $d\sigma(e^++e^-\rightarrow \chi_{c1}+
\delta_1)/dcos\theta$ & $0.0307(1+0.922cos^2\theta)$ \\\hline
  $d\sigma(e^++e^-\rightarrow \chi_{c1}+
\delta_2)/dcos\theta$ & $0.0545(1-0.273cos^2\theta)$ \\\hline
  $d\sigma(e^++e^-\rightarrow \chi_{c1}+
\delta_3)/dcos\theta$ & $0.0163(1+0.760cos^2\theta)$ \\\hline
  $d\sigma(e^++e^-\rightarrow \chi_{c2}+
\delta_1)/dcos\theta$ & $0.0161(1+0.952cos^2\theta)$ \\\hline
  $d\sigma(e^++e^-\rightarrow \chi_{c2}+
\delta_2)/dcos\theta$ & $0.00331(1+0.799cos^2\theta)$ \\\hline
  $d\sigma(e^++e^-\rightarrow \chi_{c2}+
\delta_3)/dcos\theta$ & $0.00530(1-0.188cos^2\theta)$ \\\hline
\end{tabular}
\end{table*}
\end{center}

\newpage
\begin{center}
\begin{table*}
\caption{Differential cross sections--angular distribution functions
for double charmonium production in $e^+ e^-$ annihilation at
$\sqrt{s}=10.6$~GeV with both QCD and QED contributions (see text
for the input parameters).}
\begin{tabular}{|l|l|}
  \hline
  Differential cross section & Angular distribution function (fb)\\\hline
  $d\sigma(e^++e^-\rightarrow J/\psi +
\eta_{c}(1S))/dcos\theta$ & $2.47(1+cos^2\theta)$ \\\hline
  $d\sigma(e^++e^-\rightarrow J/\psi +
\eta_{c}(2S))/dcos\theta$ & $1.62(1+cos^2\theta)$ \\\hline
  $d\sigma(e^++e^-\rightarrow J/\psi + \eta_{c}(3S))/dcos\theta$ &
$1.39(1+cos^2\theta)$ \\\hline
  $d\sigma(e^++e^-\rightarrow J/\psi +
\chi_{c0})/dcos\theta$ & $3.18(1+0.265cos^2\theta)$ \\\hline
  $d\sigma(e^++e^-\rightarrow J/\psi +
\chi_{c1})/dcos\theta$ & $0.426(1+0.751cos^2\theta)$ \\\hline
  $d\sigma(e^++e^-\rightarrow J/\psi +
\chi_{c2})/dcos\theta$ & $0.929(1-0.161cos^2\theta)$ \\\hline
  $d\sigma(e^++e^-\rightarrow J/\psi(1S)+
^1D_2)/dcos\theta$ & $0.0770(1+cos^2\theta)$ \\\hline
\end{tabular}
\end{table*}
\end{center}

\newpage
\begin{center}
\begin{table*}
\caption{Ratios of production cross sections
of various double charmonia to that of $J/\psi+\eta_c$ in $e^+ e^-$
annihilation at $\sqrt{s}=10.6$~GeV. }
\begin{tabular}{|l|l|l|l|l|l|l|l|l|l|}
\hline
~~&$\eta_{c}(1S,2S,3S)$&$\chi_{c0}(1P,2P)$&$\chi_{c1}(1P,2P)$&$\chi_{c2}(1P,2P)$&$h_{c}(1P,2P)$
&$^3D_{1}$& $^3D_{2}$&$^3D_{3}$&$^1D_2$
\\\hline
$\psi$(1S)&1.0,0.65,0.56&1.05,1.4&0.15,0.21&0.27,0.36&&&&&0.03
\\\hline
$\psi(2S)$&0.65,0.42,0.36&0.68,0.94&0.11,0.14&0.17,0.24&&&&&0.02
\\\hline $\eta_{c}$(1S)&&&&&0.11,0.15&0.025&&&
\\\hline $\eta_{c}$(2S)&&&&&0.07,0.10&0.016&&&
\\\hline $\eta_{c}$(3S)&&&&&0.06,0.08&&&&
\\\hline $\chi_0$(1P)&&&&&0.03,0.05&&&&
\\\hline $\chi_1$(1P)&&&&&0.15,0.21&0.012&0.015&0.006&
\\\hline $\psi_2$(1P)&&&&&0.010,0.013&0.006&0.001&0.0015&
\\\hline
\end{tabular}
\end{table*}
\end{center}

\newpage
\begin{figure}[h]
\begin{center}
\includegraphics[width=12cm,height=14cm]{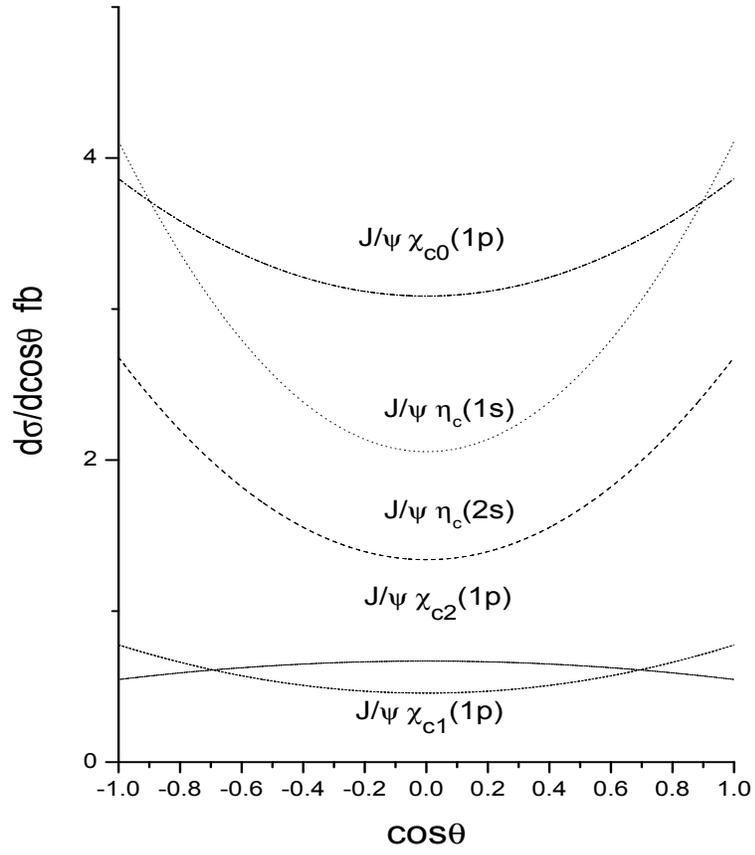}
\end{center}
\caption{ Differential cross sections for $e^+e^- \rightarrow
J/\psi+\chi_{cJ}(\eta_c)$}
\end{figure}

\newpage

\begin{figure}[h]
\begin{center}
\includegraphics[width=12cm,height=14cm]{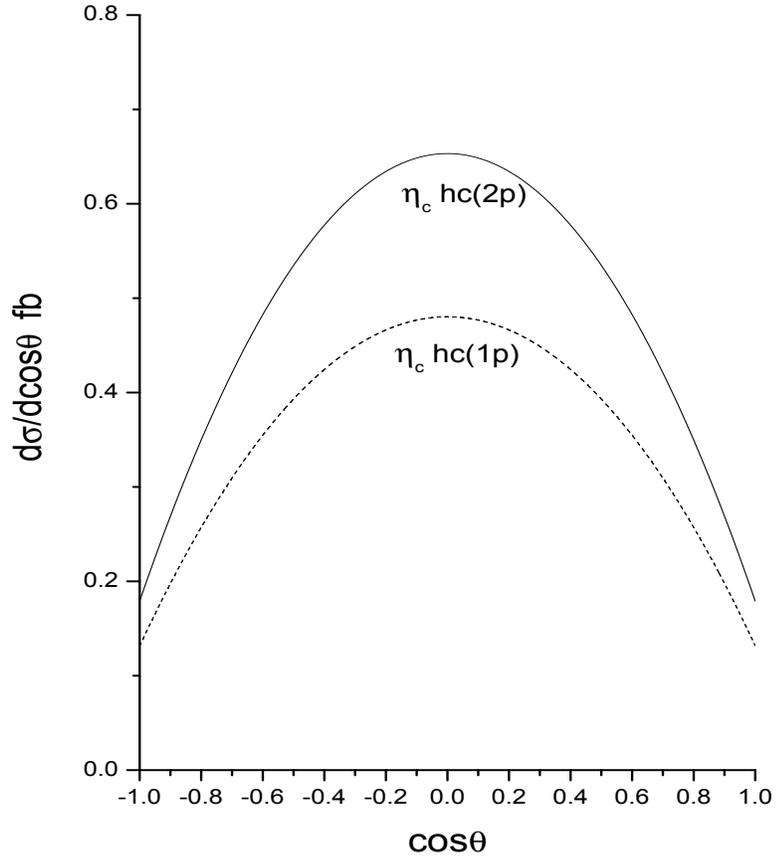}
\end{center}
\caption{ Differential cross sections for $e^+e^- \rightarrow
\eta_c+ h_c$}
\end{figure}

\newpage

\begin{figure}[h]
\begin{center}
\includegraphics[width=12cm,height=14cm]{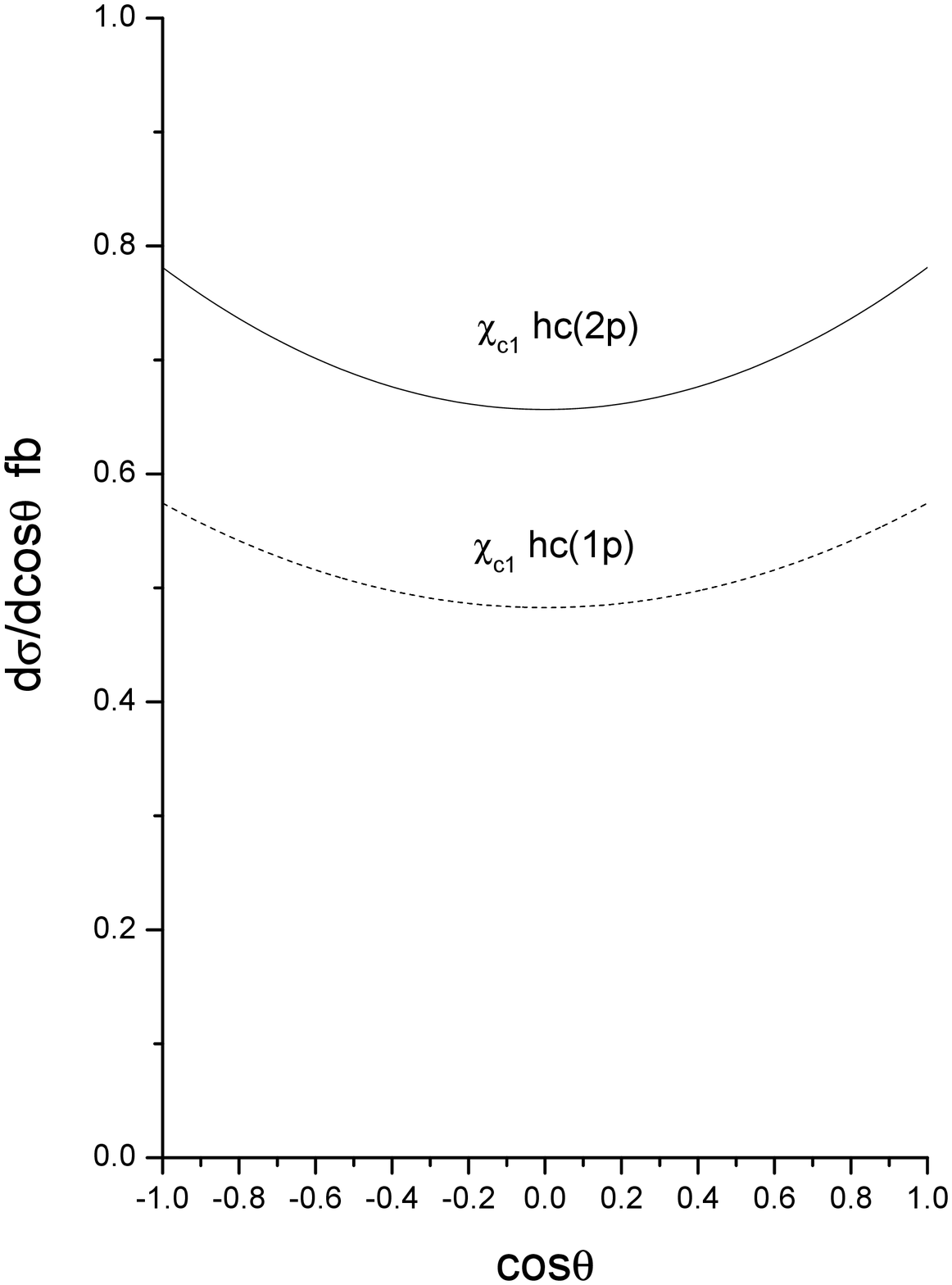}
\end{center}
\caption{ Differential cross sections for $e^+e^- \rightarrow
\chi_{c1}+h_c$}
\end{figure}

\newpage

\begin{figure}[h]
\begin{center}
\includegraphics[width=12cm,height=14cm]{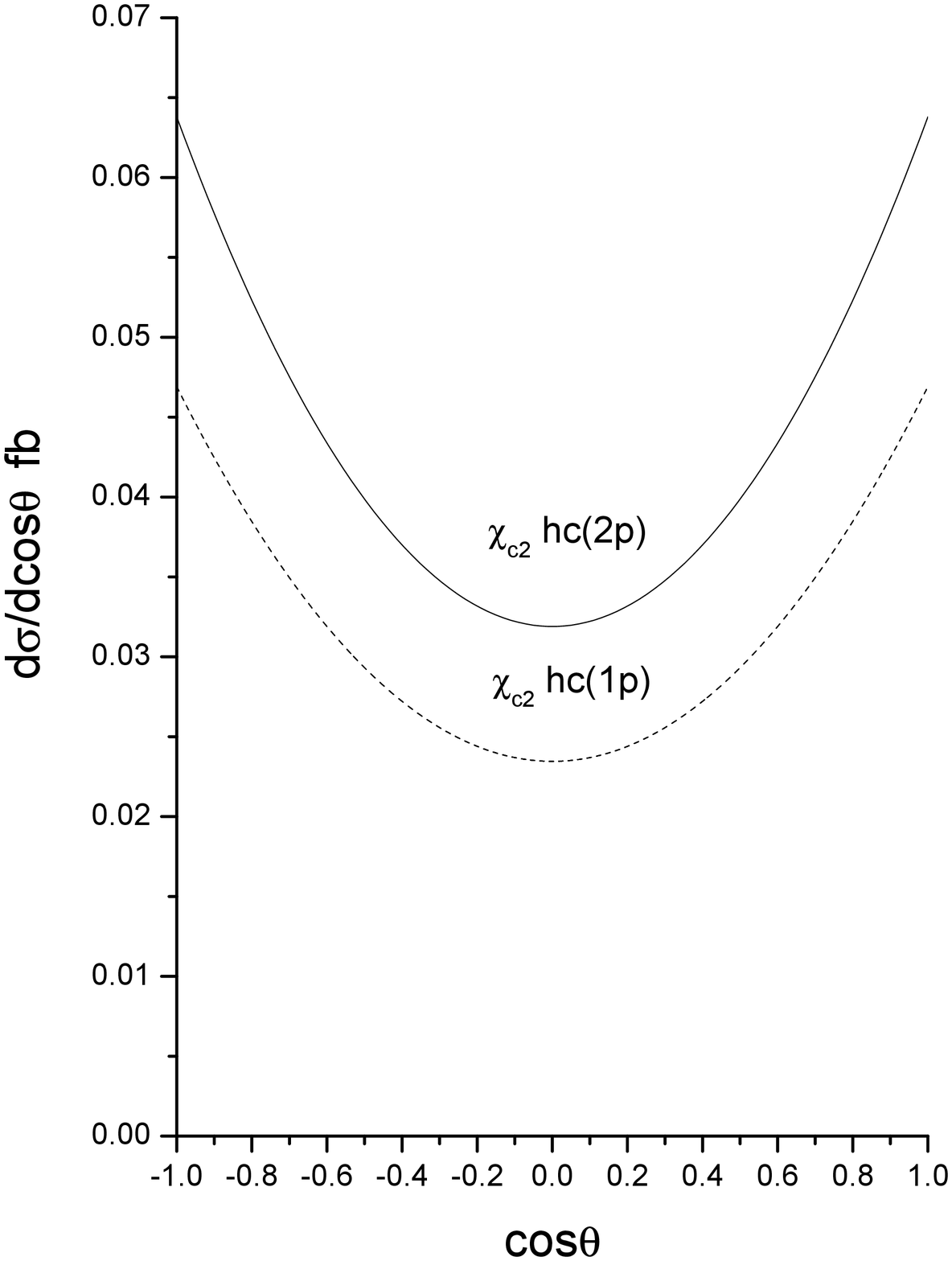}
\end{center}
\caption{ Differential cross sections for $e^+e^- \rightarrow
\chi_{c2}+ h_c$}
\end{figure}

\newpage

\begin{figure}[h]
\begin{center}
\includegraphics[width=12cm,height=14cm]{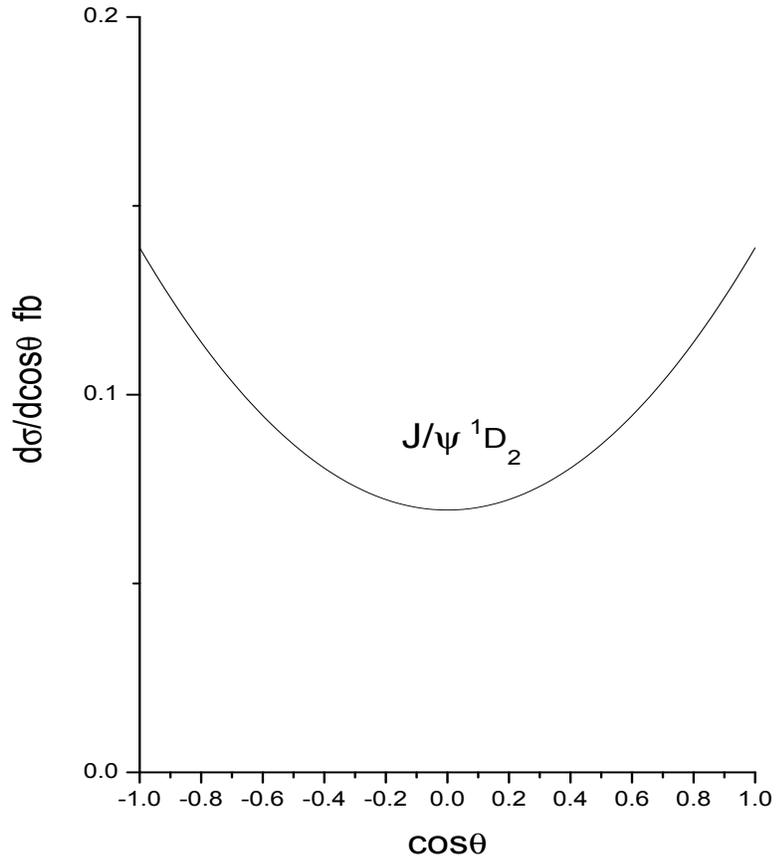}
\end{center}
\caption{ Differential cross sections for $e^+e^- \rightarrow
J/\psi+^1D_2$}
\end{figure}

\newpage

\begin{figure}[h]
\begin{center}
\includegraphics[width=12cm,height=14cm]{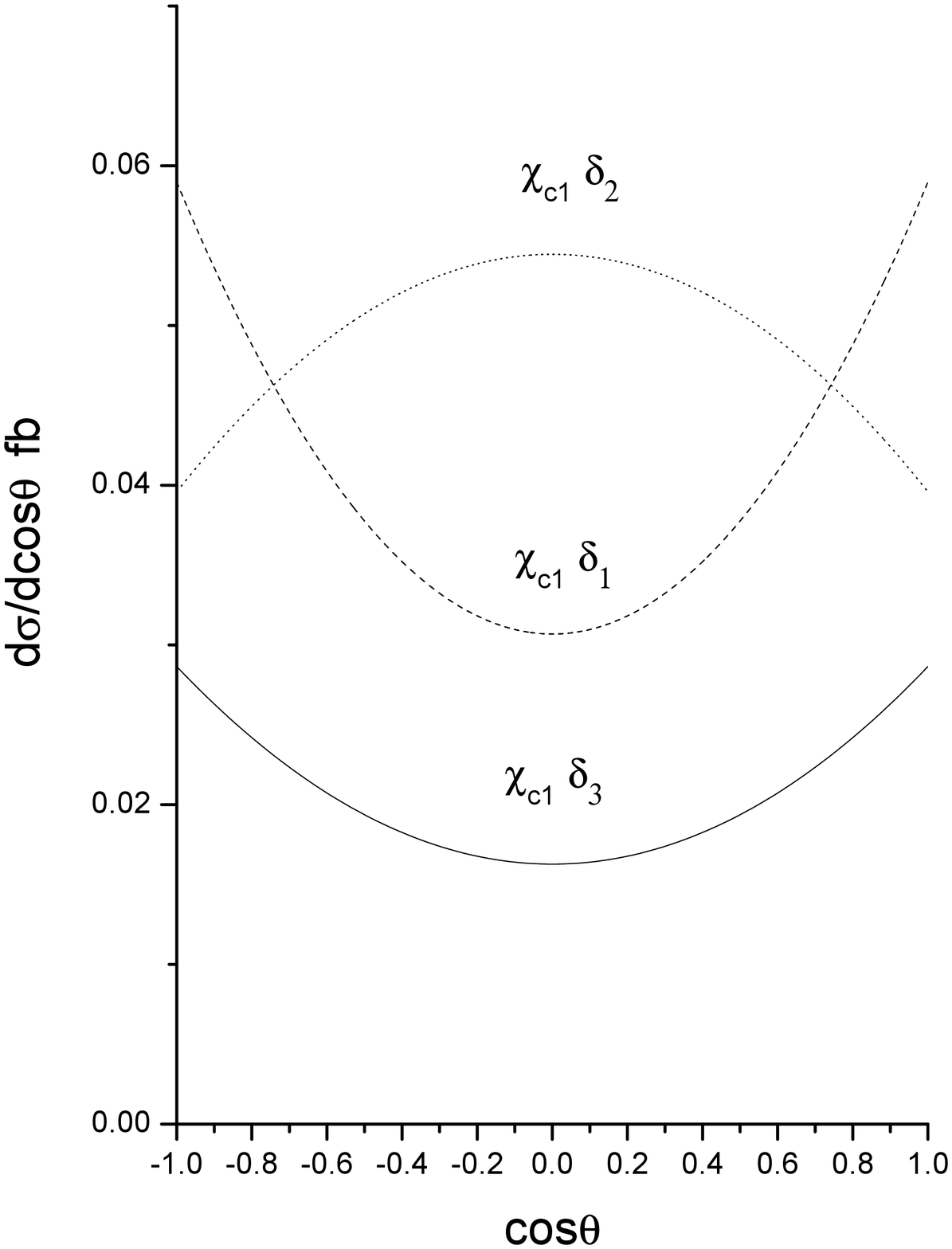}
\end{center}
\caption{ Differential cross sections for $e^+e^- \rightarrow
\chi_{c1}+\delta_J$}
\end{figure}

\newpage

\begin{figure}[h]
\begin{center}
\includegraphics[width=12cm,height=14cm]{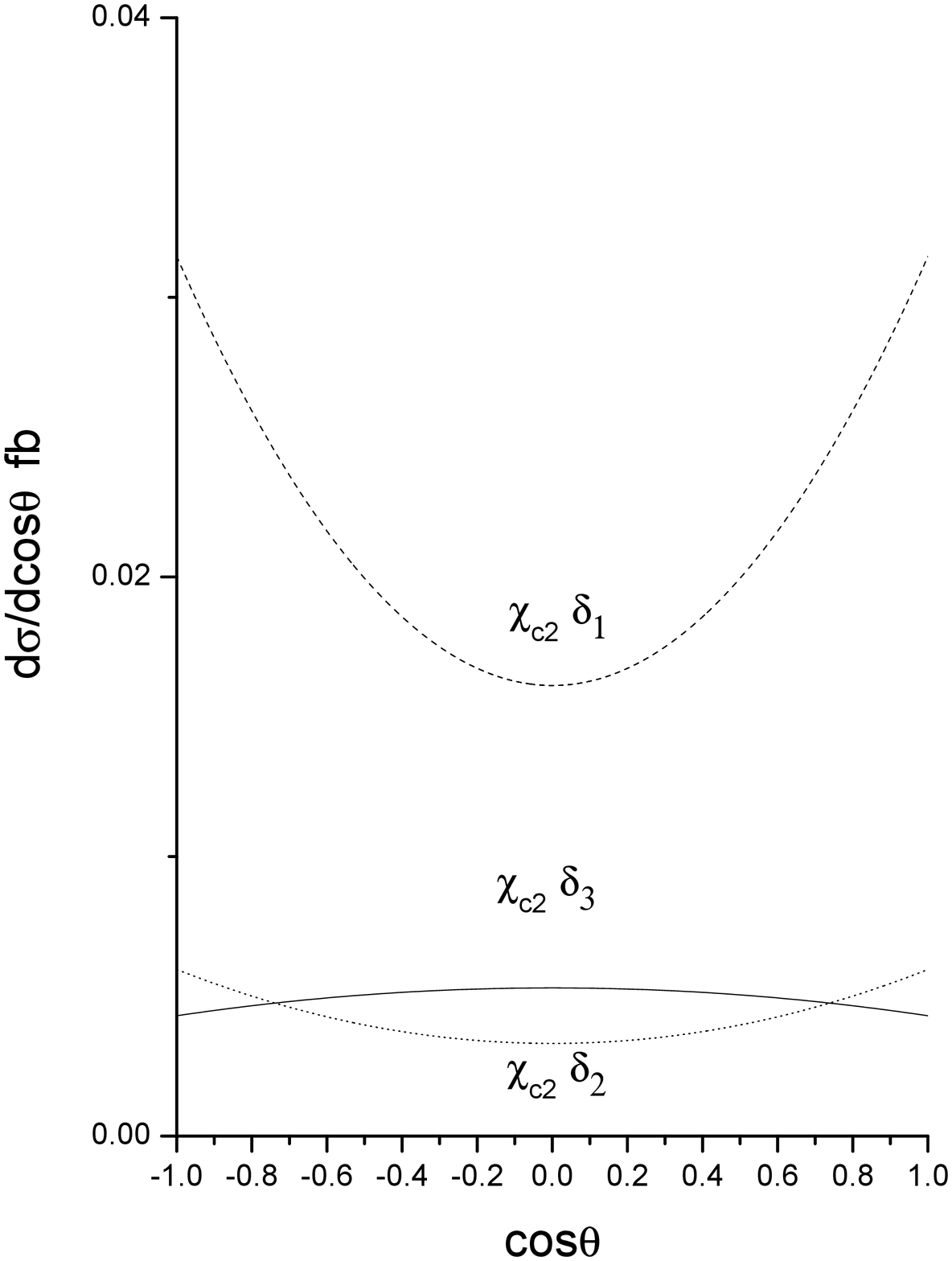}
\end{center}
\caption{ Differential cross sections for $e^+e^- \rightarrow
\chi_{c2}+\delta_J$}
\end{figure}

\newpage

\begin{figure}[h]
\begin{center}
\includegraphics[width=12cm,height=14cm]{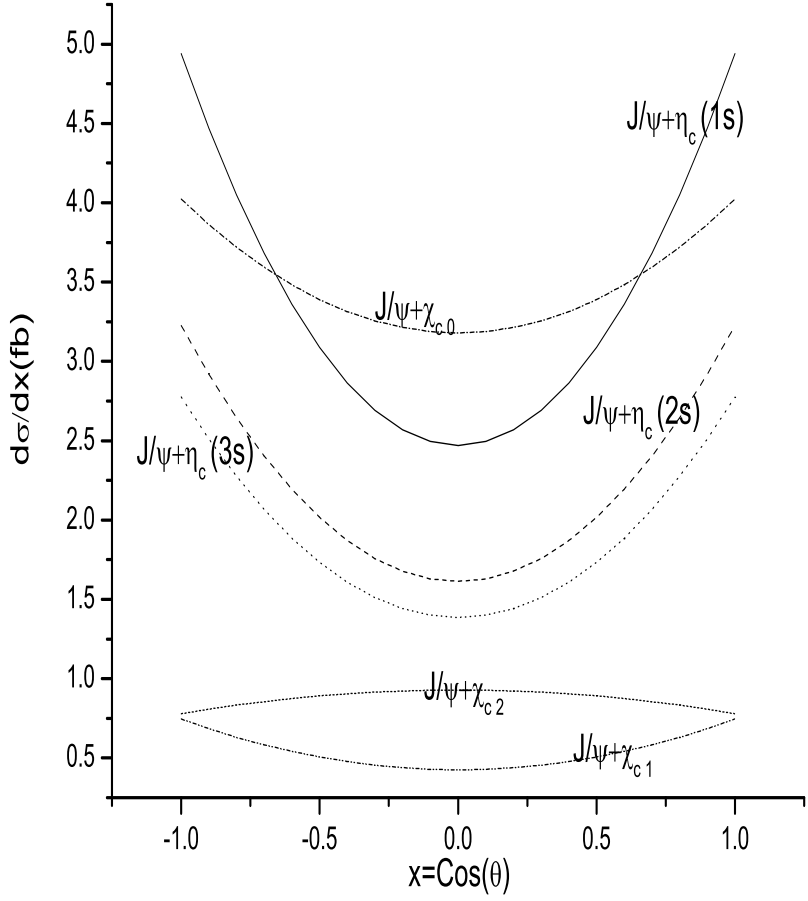}
\end{center}
\caption{ Differential cross sections with both QCD and QED
contributions for $e^+e^- \rightarrow J/\psi\eta_c(1S,2S,3S)$ and
$e^+e^- \rightarrow J/\psi+\chi_{cJ}(J=0,1,2)$}
\end{figure}

\end{document}